\documentstyle[11pt]{article}

\begin{document}

\newcommand\zero{{\tt 0}}
\newcommand\threezero{{\tt 000}}
\newcommand\one{{\tt 1}}
\newcommand\threeone{{\tt 111}}
\newcommand\plus{{\mathord{+}}}
\newcommand\minus{{\mathord{-}}}
\newcommand\up{{\mathord{\uparrow}}}
\newcommand\down{{\mathord{\downarrow}}}
\newcommand\tr{\mbox{\rm tr}}
\newcommand\Tr{\mbox{\rm Tr}}
\newcommand\cross{{\mathord{\times}}}

\title{Quantum Error Correction and Reversible Operations}

\author{Carlton M.~Caves\thanks{
I thank H.~Barnum, C.~A. Fuchs, and M.~A. Nielsen for occasional
encouragement on the superoperator formalism laid out in Sec.~\ref{supops},
and I particularly thank M.~A. Nielsen for suggesting the work outlined in 
Sec.~\ref{thebounds}.  This work was supported in part by the Office of 
Naval Research (Grant No.~N00014-93-1-0016).}
\\
\\
Center for Advanced Studies\\
 Department of Physics and Astronomy\\ 
University of New Mexico\\
Albuquerque, NM 87131\, USA}
\date{\today}
\maketitle

\begin{abstract}
I give a pedagogical account of Shor's nine-bit code for correcting
arbitrary errors on single qubits, and I review work that determines when it
is possible to maintain quantum coherence by reversing the deleterious effects
of open-system quantum dynamics.  The review provides an opportunity to
introduce an efficient formalism for handling superoperators.  I present
and prove some bounds on entanglement fidelity, which might prove useful
in analyses of approximate error correction. 
\end{abstract}

\section{Introduction}
\label{intro}

A bit is the fundamental unit of information, represented by a choice between 
two alternatives, conventionally labeled \zero\ and \one.  In the real world
the abstract notion of a bit must be realized as a physical system.  A 
{\it classical bit}, the unit of classical information processing, can be 
thought of as a two-state classical system.  The two states, perhaps \zero\ 
or \one\ printed on a page or the two positions of a particle in a double-well 
system, are distinguishable, and because they are distinguishable, they can
be copied.   A {\it quantum bit\/} or {\it qubit\/}, the unit of quantum 
information processing, is a two-state quantum system.  The two basis states, 
labeled $|\zero\rangle$ and $|\one\rangle$, are distinguishable and can be 
copied, just like the states of a classical bit.  The difference arises from 
the superposition principle: the general state of a qubit is an arbitrary 
linear combination of $|\zero\rangle$ and $|\one\rangle$.  The many possible 
superposition states available to a qubit give it more information-processing 
power than a classical bit, even though the general superposition states 
cannot be distinguished reliably and cannot be copied---the no-cloning theorem
forbids \cite{Wootters1982a,Dieks1982a}.  The enhanced information-processing 
power of qubits can be harnessed to a variety of information-processing tasks 
including computation (for a review and extensive list of references, see 
\cite{Steane1998a}).  

The price for the power of quantum information is eternal vigilance in 
maintaining quantum coherence, for the enhanced information-processing power 
comes from the ability to manipulate superposition states.  Coupling to the 
environment, with the accompanying noise and decoherence, tends to destroy 
superpositions.  Indeed, what we call a classical bit is just a qubit whose 
coupling to the environment keeps it from occupying superpositions of two 
orthogonal states.  Decoherence strips the kets off $|\zero\rangle$ and 
$|\one\rangle$, leaving the \zero\ and \one\ of a classical bit, which 
cannot be superposed.

Quantum information processing got a tremendous boost with Shor's surprising
discovery \cite{Shor1995a} that quantum information stored in superpositions 
can be protected against decoherence. Shor's announcement of a nine-bit quantum
code, followed shortly by the discovery of five-bit \cite{Laflamme1996a}
and seven-bit codes \cite{Steane1996a}, ignited an explosion of activity on 
quantum error correction 
\cite{Steane1996b,Calderbank1996a,Gottesman1996a,Schumacher1996b,Bennett1996b,%
Knill1997a} (see \cite{Steane1998a} for further references), which led to the 
demonstration that a quantum computer can perform arbitrarily complicated 
computations provided that the error per operation can be reduced below an 
error threshold \cite{Gottesman1998a,Preskill1998a,Knill1998b}.  

In this article I give in Sec.~\ref{errcorr} a pedagogical account of the 
simplest error-correction scheme, Shor's nine-bit code \cite{Shor1995a}, 
with the aim of illustrating the essential ideas of quantum error correction.  
In Sec.~\ref{qops} I review work that determines when it is possible to 
maintain quantum coherence by reversing the deleterious effects of 
open-system quantum dynamics, using the review as an opportunity to introduce
an efficient formalism for manipulating superoperators.  In Sec.~\ref{bounds} 
I first review the information-theoretic description of error correction and 
then present and prove some bounds on entanglement fidelity, which might 
turn out to be useful in considering approximate error correction.

\section{Quantum Error Correction: Shor's Nine-Bit Code}
\label{errcorr}

Quantum error correction is closely related to classical error correction.  
Much of the theory of quantum error correction 
\cite{Steane1996b,Calderbank1996a,Gottesman1996a} comes from the theory of 
classical linear codes, in which a classical code word of length $k$, i.e., 
a string of $k$ \zero's and \one's, becomes a vector in a $k$-dimensional 
vector space over the field consisting of 0 and 1.  I do not discuss this 
theory here, but rather consider a particular quantum code, Shor's nine-bit 
code, to illustrate the essential ideas of quantum error correction.

To get started, though, let's first consider correcting errors on a classical 
bit.  The only type of error is a bit flip, in which \zero\ and \one\ are 
interchanged. If the errors are rare, they can be corrected using redundancy, 
because one only needs to correct single-bit errors.  Suppose, for example, 
that a \zero\ is encoded as the three-bit string \threezero, and a \one\ as 
the three-bit string \threeone.  Then an error on a single bit can always be 
detected and corrected.  The situation is summarized here:
\begin{equation}
\matrix{
\mbox{\rm no error}\qquad&\threezero\quad&\threeone\;,\cr 
\mbox{\rm flip 1st bit}\qquad&{\tt100}\quad&{\tt011}\;,\cr 
\mbox{\rm flip 2nd bit}\qquad&{\tt010}\quad&{\tt101}\;,\cr 
\mbox{\rm flip 3rd bit}\qquad&{\tt001}\quad&{\tt110}\;.
}
\end{equation}
An error can be detected by polling the three bits.  The minority bit 
suffered the flip, which can be corrected by flipping the minority bit
again.  If the probability for a bit flip on a single qubit is $p\ll1$, 
then this scheme reduces the probability of error from $p$ to $3p^2$, 
a winning proposition if $p\le{1\over3}$.

The key elements in this scheme, redundancy and polling, cannot be 
translated to quantum error correction: making redundant copies of a qubit 
is ruled out by the no-cloning theorem, which forbids making copies of 
arbitrary superposition states; polling requires ascertaining the state 
of each qubit, which destroys the quantum coherence one is aiming to protect.  
On closer examination, though, the situation is more promising.  The eight 
strings that result from no error and from the three bit flips are all 
distinct.  This distinguishability opens up the possibility of a 
{\it quantum code\/} in which $|\zero\rangle$ is encoded as a state of three 
qubits, the ``logical zero'' state $|\zero\rangle_L=|\threezero\rangle$, and 
$|\one\rangle$ is encoded as the ``logical one'' state 
$|\one\rangle_L=|\threeone\rangle$.  The logical zero and logical one span 
a two-dimensional subspace, the ``code subspace'' to be used for quantum 
information processing.  Notice now what happens to an arbitrary superposition 
of the logical one and logical zero states under single-qubit bit flips:
\begin{equation}
\matrix{
\mbox{\rm no error}\;&1\otimes1\otimes1\;&
\alpha|\threezero\rangle+\beta|\threeone\rangle\;,\cr 
\mbox{\rm flip 1st qubit}\;&\sigma_1\otimes1\otimes1\;&
\alpha|{\tt100}\rangle+\beta|{\tt011}\rangle\;,\cr 
\mbox{\rm flip 2nd qubit}\;&1\otimes\sigma_1\otimes1\;&
\alpha|{\tt010}\rangle+\beta|{\tt101}\rangle\;,\cr 
\mbox{\rm flip 3rd qubit}\;&1\otimes1\otimes\sigma_1\;&
\alpha|{\tt001}\rangle+\beta|{\tt110}\rangle\;.
}
\end{equation}
Here the middle column writes the error in terms of the unit operator
$1$ and the bit-flip Pauli operator 
$\sigma_1=|\zero\rangle\langle\one|+|\one\rangle\langle\zero|$ for the 
appropriate qubit.  

The no-error operator and the three bit-flip errors map the code subspace 
{\it unitarily\/} to {\it orthogonal\/} two-dimensional subspaces within 
the eight-dimensional Hilbert space of a qubit triplet.  The single-qubit 
errors can be detected and distinguished by determining in which 
two-dimensional subspace the system lies, without in any way disturbing the 
superposition, and the error can be corrected by mapping the error subspaces 
unitarily back to the code subspace.  This is the essence of quantum error 
correction: find a code subspace such that the high probability errors map 
unitarily to orthogonal subspaces; then the errors can be detected and 
corrected without destroying quantum coherence.  The superposition states 
in the code subspace are entangled states of the three qubits.  The redundancy 
used by a classical code, which cannot be translated to quantum coding, is 
replaced by {\it entanglement\/} in a quantum code.  

Having gotten this far, however, we now realize that the task is tougher than 
classical error correction, because there are quantum errors that have no 
classical counterpart.  Specifically, there are ``phase flips,'' described 
by the Pauli operator 
$\sigma_3=|\zero\rangle\langle\zero|-|\one\rangle\langle\one|$, and errors 
described by the Pauli operator $-i\sigma_2=\sigma_1\sigma_3=
-|\zero\rangle\langle\one|+|\one\rangle\langle\zero|$, which is a phase flip 
followed by a bit flip.  If we can correct these errors, we can correct
all single-qubit errors, because all error operators can be written as a
linear combination of the unit operator (no error) and the three Pauli
operators.  

Let's concentrate first on the phase-flip errors, hoping that the combined
phase-bit flips take care of themselves.  The first thing to notice is that 
$\sigma_1$ and $\sigma_3$ switch roles in the transformed basis defined by
\begin{equation}
|\pm\rangle\equiv{1\over\sqrt2}\bigl(|\zero\rangle\pm|\one\rangle\bigr)\;:
\end{equation}
$\sigma_3$ becomes a bit flip, and $\sigma_1$ becomes a phase flip,
\begin{equation}
\sigma_3|\pm\rangle=|\mp\rangle\;,\qquad
\sigma_1|\pm\rangle=\pm|\pm\rangle\;.
\end{equation}
Thus we can correct single-qubit phase flips by using a quantum code whose 
logical basis states are $|\plus\plus\plus\rangle$ and 
$|\minus\minus\minus\rangle$, but this comes at the expense of being unable
to correct the original bit-flip errors.  

Here entanglement comes to the rescue again.  The entire code subspace 
spanned by $|\threezero\rangle$ and $|\threeone\rangle$ is protected against 
single-qubit bit flips, so we can use any orthogonal basis in the code subspace 
as the logical zero and one.  In particular, we can use ``up'' and ``down'' 
states 
\begin{equation}
\matrix{
|\up\rangle
\equiv{1\over\sqrt2}\bigl(|\threezero\rangle+|\threeone\rangle\bigr)\;,\cr
|\down\rangle
\equiv{1\over\sqrt2}\bigl(|\threezero\rangle-|\threeone\rangle\bigr)\;,}
\end{equation}
which are defined in analogy to the way the states $|\pm\rangle$ are related 
to $|\zero\rangle$ and $|\one\rangle$ for a single qubit.  The effect of
single-qubit bit flips on these states is summarized here:
\begin{equation}
\matrix{
\sigma_1\otimes1\otimes1|\up\rangle=
{1\over\sqrt2}
\bigl(|\one\zero\zero\rangle+|\zero\one\one\rangle\bigr)
\equiv|\up_1\rangle\;,\cr
1\otimes\sigma_1\otimes1|\up\rangle=
{1\over\sqrt2}
\bigl(|\zero\one\zero\rangle+|\one\zero\one\rangle\bigr)
\equiv|\up_2\rangle\;,\cr
1\otimes1\otimes\sigma_1|\up\rangle=
{1\over\sqrt2}
\bigl(|\zero\zero\one\rangle+|\one\one\zero\rangle\bigr)
\equiv|\up_3\rangle\;,\cr
\vphantom{\scriptscriptstyle{x}}\cr
\sigma_1\otimes1\otimes1|\down\rangle=
{1\over\sqrt2}
\bigl(|\one\zero\zero\rangle-|\zero\one\one\rangle\bigr)
\equiv|\down_1\rangle\;,\cr
1\otimes\sigma_1\otimes1|\down\rangle=
{1\over\sqrt2}
\bigl(|\zero\one\zero\rangle-|\one\zero\one\rangle\bigr)
\equiv|\down_2\rangle\;,\cr
1\otimes1\otimes\sigma_1|\down\rangle=
{1\over\sqrt2}
\bigl(|\zero\zero\one\rangle-|\one\one\zero\rangle\bigr)
\equiv|\down_3\rangle\;.
}
\end{equation}
The eight types of up and down states make up an orthogonal basis for the 
three-qubit Hilbert space.

Notice now that the bare up and down states are flipped by all three 
single-qubit phase flips:
\begin{equation}
\left(
\matrix{
\sigma_3\otimes1\otimes1\cr 
\mbox{or}\cr
1\otimes\sigma_3\otimes1\cr 
\mbox{or}\cr
1\otimes1\otimes\sigma_3
} 
\right)
\times
\left\{
\matrix{|\up\rangle\cr|\down\rangle}
\right\}
=
\left\{
\matrix{|\down\rangle\cr|\up\rangle}
\right\}\;.
\end{equation}
This suggests correcting both bit- and phase-flip errors by again tripling 
the number of qubits and using the following nine-qubit states as the 
logical zero and one states:
\begin{equation}
|\zero\rangle_L=|\up\up\up\rangle\;,\quad
|\one\rangle_L=|\down\down\down\rangle\;.
\end{equation}
This logical zero and one constitute Shor's nine-bit code.  

To complete the discussion of the nine-bit code, I display in
Table~\ref{table1} how an arbitrary state in the code subspace is affected 
by all 27 single-qubit errors.  Notice first that in accordance with the 
discussion above, the three phase-flip errors on each qubit triplet have 
exactly the same effect.  Thus the nine-bit code is a {\it degenerate code}
\cite{Gottesman1996a}, one in which errors that are independent on the 
entire Hilbert space become the same on the code subspace.  The degeneracy 
leaves a situation where we must consider the no-error case plus 21 
independent errors.  Each error maps the code subspace unitarily to a 
two-dimensional subspace. Moreover, by examining the table, one sees that 
the code subspace and the 21 error subspaces are mutually orthogonal.
The 22 states that arise from $|\zero\rangle_L$ are orthogonal to the 22 
states that arise from $|\one\rangle_L$: the former all have up states for 
two of the triplets, whereas the latter all have down states for two of 
triplets, so they disagree on up versus down in at least one position.  
The states that arise from $|\zero\rangle_L$ ($|\one\rangle_L$) are mutually 
orthogonal because they are the 22 states that come from putting two of 
the triplets in the up (down) state, with the third triplet cycling through 
the eight up- and down-type states that span the triplet Hilbert space.  
Single-qubit errors can be corrected by determining in which of the 
orthogonal subspaces the nine qubits lie and then mapping that subspace 
unitarily back to the code subspace.

As noted above, error correction works for single-qubit errors that are 
not described by Pauli operators, because any error can be written as a
linear combination of the unit operator and the Pauli operators.  Consider, 
for example, decay from $|\zero\rangle$ to $|\one\rangle$, with essentially
instantaneous phase decoherence between $|\zero\rangle$ and $|\one\rangle$, 
a situation described by three error operators: 
$A_1=\sqrt{1-\gamma}|\zero\rangle\langle\zero|=\sqrt{1-\gamma}(1+\sigma_3)/2$,
$A_2=\sqrt{\gamma}|\one\rangle\langle\zero|=
\sqrt{\gamma}(\sigma_1-i\sigma_2)/2$,
and $A_3=|\one\rangle\langle\one|=(1-\sigma_3)/2$, $\gamma$ being the
probability of decay.  If one of the nine qubits suffers a decay, a 
measurement of the error subspace reveals either no error or one of the
Pauli errors on that qubit.  All of these being correctable, it doesn't
matter which is the result of the measurement.  

The nine-bit code wastes Hilbert space, for it uses only 44 of the 
$2^9=512$ dimensions in the nine-qubit Hilbert space.  If one wants to correct 
$r$ errors per qubit, using a code with $N$ qubits, then to accommodate the 
code subspace and the $rN$ errors, one needs $2(1+rN)$ dimensions.  This 
leads to the {\it quantum Hamming bound} \cite{Gottesman1996a}:
\begin{equation}
2(1+rN)\le2^N\quad\Longleftrightarrow\quad r\le{2^{N-1}-1\over N}\;.
\end{equation}
Three qubits permit correction of one error, the situation we started with 
in this section.  Five qubits have the potential for correcting the three 
Pauli errors, a potential realized in an astonishing five-qubit quantum 
code \cite{Laflamme1996a}.

\section{Reversal of Open-System Dynamics}
\label{qops}

In this section I review the description of open-system dynamics in terms
of quantum operations and the question of when quantum coherence can be 
maintained by reversing open-system dynamics.  The review serves to introduce 
a formalism for handling quantum operations, which provides insight into 
their structure.  

Throughout this section and the next I use a set of conventions introduced 
by Schumacher \cite{Schumacher1996a}.  The {\it primary quantum system\/} is 
denoted by $Q$; it is assumed to be finite-dimensional, with a Hilbert space
${\cal H}_Q$ of dimension $D$.  The primary system interacts with an 
{\it environment\/} $E$, and to deal with purifications of $Q$ states, 
there can be a additional, passive {\it reference system\/} $R$.  Where it 
is necessary to avoid confusion, superscripts $R$, $Q$, and $E$ are used to 
distinguish states and operators of these systems.  Initial states are 
unprimed, and states after the dynamics are denoted by a prime.

\subsection{Open-system dynamics and quantum operations}

Consider a primary system $Q$, initially in state $\rho$, which is brought 
into contact with an environment $E$, initially in state 
$\rho^E=\sum_l\lambda_l|\phi_l\rangle\langle\phi_l|$, where the states 
$|\phi_l\rangle$ are the eigenstates of $\rho^E$.  The two systems interact 
for a time, the interaction described by a unitary operator $U$, and then 
the environment is observed to be in a subspace spanned by orthogonal states 
$|g_j\rangle$, corresponding to a projection operator
$P^E=\sum_k|g_k\rangle\langle g_k|$.  The unnormalized state of the system 
after the observation is given by a partial trace over the environment,
\begin{equation}
\tr_E\bigl(P^E\,U(\rho\otimes\rho^E)U^\dagger\bigr)
\equiv{\cal A}(\rho)\;,
\label{poststate}
\end{equation}
where ${\cal A}$ is a linear map on system density operators.  Inserting 
the forms of the projector $P^E$ and the environment state $\rho^E$ leads
to
\begin{equation}
{\cal A}(\rho)=
\sum_{k,l}
\sqrt{\lambda_l}\langle g_k|U|\phi_l\rangle
\rho
\langle\phi_l|U^\dagger|g_k\rangle\sqrt{\lambda_l}
=\sum_\alpha A_\alpha\rho A_\alpha^\dagger\;.
\label{opdecomp}
\end{equation}
The system operators
\begin{equation}
A_\alpha=A_{kl}
\equiv
\sqrt{\lambda_l}\langle g_k|U|\phi_l\rangle\;,
\label{decompops}
\end{equation}
where the Greek index $\alpha$ is as an abbreviation for $k$ and $l$, 
provide an {\it operator decomposition\/} of the map ${\cal A}$ and 
thus are called {\it decomposition operators}.  The {\it normalized\/} 
post-dynamics system state is 
\begin{equation}
\rho'={{\cal A}(\rho)\over\tr\bigl({\cal A}(\rho)\bigr)}\;,
\end{equation}
where one should notice that 
\begin{equation}
\tr\bigl(P^E\,U(\rho\otimes\rho^E)U^\dagger\bigr)=
\tr\bigl({\cal A}(\rho)\bigr)=
\tr\biggl(\rho
\sum_\alpha A_\alpha^\dagger A_\alpha
\biggr)
\label{Eprob}
\end{equation}
is the probability for the environment to be found in the specified subspace.

A primary quantum system that is exposed to an initially uncorrelated 
environment {\it always\/} has dynamics described by a map like ${\cal A}$.  
This includes both {\it nonselective dynamics}, where no observation is made 
on the environment ($P^E=1^E$), and {\it selective dynamics}, where the 
system state is conditioned on the result of a measurement on the environment 
($P^E\ne1^E$).  It is important to identify the mathematical conditions that 
characterize a suitable map ${\cal A}$.  We can immediately identify three 
conditions that ${\cal A}$ must satisfy.
\vskip 2pt
{\it Condition~1}.  ${\cal A}$ is a {\it linear\/} map on operators; i.e., 
it is a {\it superoperator}.
\vskip 2pt
{\it Condition~$2'$}.  ${\cal A}$ maps positive operators to positive 
operators. (A positive operator $G$ is one such that
$\langle\psi|G|\psi\rangle\ge0$ for all vectors $|\psi\rangle$; density 
operators are positive operators.)
\vskip 2pt
{\it Condition~3}.  ${\cal A}$ is {\it trace-decreasing}, i.e., 
$\tr\bigl({\cal A}(\rho)\bigr)\le1$ for all density operators $\rho$.  This
condition, which follows immediately from Eq.~(\ref{Eprob}), can be expressed
as an operator inequality, 
\begin{equation}
\sum_\alpha A_\alpha^\dagger A_\alpha\le 1\;.
\label{tdcond}
\end{equation}
The trace is preserved for nonselective dynamics, but generally decreases
for selective dynamics.
\vskip2pt
It might be thought that the above three conditions are sufficient to 
characterize ${\cal A}$, but it turns out that Condition~$2'$ must be
strengthened.  The more restrictive condition can be motivated physically.
Suppose that $R$ is a reference system that, though it does not take part 
in the dynamics, cannot be neglected because the initial state $\rho$ 
of $Q$ is the partial trace over $R$ of a joint state $\rho^{RQ}$.  We 
certainly want the map ${\cal I}^R\otimes{\cal A}$, where ${\cal I}^R$ is 
the unit superoperator on $R$, to take $\rho^{RQ}$ to a positive operator, 
which can be normalized to be a density operator.  This condition holds 
trivially for a map of the form~(\ref{opdecomp}),
\begin{equation}
\bigl({\cal I}^R\otimes{\cal A}\bigr)\bigl(\rho^{RQ}\bigr)=
\sum_\alpha 
(1^R\otimes A_\alpha)\rho^{RQ}(1^R\otimes A_\alpha^\dagger)
\ge 0\;,
\end{equation}
so we replace Condition~$2'$ with a stronger condition. 
\vskip2pt
{\it Condition~2}.  ${\cal A}$ is {\it completely positive}; i.e., 
$\bigl({\cal I}^R\otimes{\cal A}\bigr)\bigl(\rho^{RQ}\bigr)\ge0$ for all 
joint density operators $\rho^{RQ}$ of $Q$ and arbitrary reference systems 
$R$.
\vskip2pt

A map on operators that satisfies Conditions~1--3 is called a {\it quantum
operation}.  The description of open-system dynamics in terms of quantum 
operations was pioneered by Hellwig and Kraus 
\cite{Hellwig1969a,Hellwig1970a,Kraus1983a}.  Thus far our discussion has
established that any open-system dynamics of the sort introduced above is 
described by a quantum operation.  We need two further properties: first, 
that any quantum operation has an operator decomposition, as in 
Eq.~(\ref{opdecomp}), and second, that any quantum operation can be 
realized by a unitary coupling to an environment.   Once the former 
result is established, the latter is easy.  Any linear, trace-decreasing 
map that has an operator decomposition can be realized by a unitary coupling 
to an {\it initially pure-state\/} environment.  One partially defines a
joint operator $U$ on $RQ$ as in Eq.~(\ref{decompops}) and uses the 
trace-decreasing condition~(\ref{tdcond}) to show that the definition can
be extended so that $U$ is unitary.  We turn now to the connection between
complete positivity and the existence of an operator decomposition, which
requires us to step back and consider the properties of superoperators.

\subsection{Superoperators and complete positivity}
\label{supops}

The space of linear operators acting on ${\cal H}_Q$ is a $D^2$-dimensional 
complex vector space ${\cal L}({\cal H}_Q)$.  Let us introduce operator 
``kets'' $|A)=A$ and ``bras'' $(A|=A^\dagger$, distinguished from vector 
kets and bras by the use of smooth brackets.  Then the natural inner product 
on ${\cal L}({\cal H}_Q)$, the trace-norm inner product, can be written as 
$(A|B)=\tr(A^\dagger B)$.  An orthonormal basis $|e_j\rangle$ induces an 
orthonormal operator basis $|e_j\rangle\langle e_k|=\tau_{jk}=\tau_\alpha$, 
where the Greek index is again an abbreviation for two Roman indices.  Not 
all orthonormal operator bases are of this outer-product form.

The space of superoperators on $Q$, i.e., linear maps on operators, is 
a $D^4$-dimensional complex vector space 
${\cal L}\bigl({\cal L}({\cal H}_Q)\bigr)$.  Any superoperator ${\cal S}$ 
is specified by its ``matrix elements''  
\begin{equation}
{\cal S}_{lj,mk}
\equiv
\bigl\langle e_l\bigl|\,
{\cal S}\bigl(|e_j\rangle\langle e_k|\bigr)
\bigr|e_m\bigr\rangle\;,
\end{equation}
for the superoperator can be written in terms of its matrix elements as
\begin{equation}  
{\cal S}=\sum_{lj,mk}
{\cal S}_{lj,mk}
|e_l\rangle\langle e_j|\otimes|e_k\rangle\langle e_m|
=
\sum_{\alpha,\beta}
{\cal S}_{\alpha\beta}\,\tau_\alpha\otimes\tau_\beta^\dagger=
\sum_{\alpha,\beta}
{\cal S}_{\alpha\beta}
|\tau_\alpha)(\tau_\beta|\;.
\end{equation}
The {\it ordinary action\/} of ${\cal S}$ on an operator $A$, used above to 
generate the matrix elements, is obtained by dropping an operator $A$ 
into the center of the representation of ${\cal S}$, in place of the 
tensor-product sign,
\begin{equation}
{\cal S}(A)=
\sum_{\alpha,\beta}
{\cal S}_{\alpha\beta}\,\tau_\alpha A\tau_\beta^\dagger\;.
\end{equation}
There is clearly another way that ${\cal S}$ can act on $A$, the 
{\it left-right action}, 
\begin{equation}
{\cal S}|A)\equiv
\sum_{\alpha,\beta}
{\cal S}_{\alpha\beta}
|\tau_\alpha)(\tau_\beta|A)\;,
\end{equation}
in terms of which the matrix elements are 
\begin{equation}
{\cal S}_{\alpha\beta}
=(\tau_\alpha|\,{\cal S}|\tau_\beta)
=(\tau_{lj}|\,{\cal S}|\tau_{mk})
=\bigl\langle e_l\bigl|\,
{\cal S}\bigl(|e_j\rangle\langle e_k|\bigr)
\bigr|e_m\bigr\rangle
\;.
\label{matel}
\end{equation}
This expression provides the fundamental connection between the two 
actions of a superoperator.

With respect to the left-right action, a superoperator works just like an
operator.  Multiplication of superoperators ${\cal T}$ and ${\cal S}$
is given by 
\begin{equation}
{\cal T\cal S}=
\sum_{\alpha,\beta,\gamma}
{\cal T}_{\alpha\gamma}{\cal S}_{\gamma\beta}
|\tau_\alpha)(\tau_\beta|\;,
\end{equation}
and the adjoint is defined by 
\begin{equation}
(A|{\cal S}^\dagger|B)=(B|{\cal S}|A)^*
\quad\Longleftrightarrow\quad
{\cal S}^\dagger=
\sum_{\alpha,\beta}
{\cal S}_{\beta\alpha}^*
|\tau_\alpha)(\tau_\beta|\;.
\end{equation}
With respect to the ordinary action, superoperator multiplication, denoted 
as a composition ${\cal T}\circ{\cal S}$, is given by
\begin{equation}
{\cal T}\circ{\cal S}=
\sum_{\alpha,\beta,\gamma,\delta}
{\cal T}_{\gamma\delta}{\cal S}_{\alpha\beta}\,
\tau_\gamma\tau_\alpha
\otimes\tau_\beta^\dagger\tau_\delta^\dagger\;.
\end{equation}
The adjoint with respect to the ordinary action, denoted by 
${\cal S}^\cross$, is defined by
\begin{equation}
\tr\bigl([{\cal S}^\cross(B)]^\dagger A\bigr)=
\tr\bigl(B^\dagger{\cal S}(A)\bigr)
\quad\Longleftrightarrow\quad
{\cal S}^\cross=
\sum_{\alpha,\beta}
{\cal S}_{\alpha\beta}^*\,
\tau_\alpha^\dagger\otimes\tau_\beta\;.
\end{equation}

To deal with complete positivity, we need to introduce a reference system $R$, 
which we choose now and henceforth to have the same dimension as $Q$, and 
we need to have a way of turning operators (superoperators) on $Q$ into 
vectors (operators) on $RQ$.  To do so, introduce the unnormalized maximally 
entangled state
\begin{equation}
|\Psi\rangle\equiv
\sum_j 
|f_j\rangle\otimes|e_j\rangle
=\sum_j |f_j,e_j\rangle\;,
\end{equation}
where the vectors $|f_j\rangle$ comprise an orthonormal basis for $R$.  The 
VEC of an operator $A$ on $Q$ is the vector 
\begin{equation}
|\Phi_A\rangle\equiv
1^R\otimes A|\Psi\rangle=
\sum_j|f_j\rangle\otimes A|e_j\rangle\;.
\end{equation}
The VEC map is a one-to-one, linear map from ${\cal L}({\cal H}_Q)$ to 
${\cal H}_{RQ}$; we recover $A$ from $|\Phi_A\rangle$ via
\begin{equation}
\langle f_j,e_k|\Phi_A\rangle=
\langle e_k|A|e_j\rangle\;.
\end{equation}
It is easy to see that VEC preserves inner products,
\begin{equation}
\langle\Phi_A|\Phi_B\rangle=
\tr(A^\dagger B)=(A|B)\;.
\end{equation}
One further aspect of VEC deserves mention.  Given a density operator $\rho$ 
for $Q$, applying VEC to $\sqrt\rho$, 
\begin{equation}
|\Phi_{\!\sqrt\rho}\rangle
=1^R\otimes\!\sqrt\rho\,|\Psi\rangle
=\sum_j|f_j\rangle\otimes\sqrt\rho\,|e_j\rangle\;,
\label{vecpure}
\end{equation}
generates a purification of $\rho$, i.e.,
\begin{equation}
\tr_R\bigl(|\Phi_{\!\sqrt\rho}\rangle\langle\Phi_{\!\sqrt\rho}|\bigr)
=\rho\;.
\end{equation}

The analogous OP map is a one-to-one, linear map from 
${\cal L}\bigl({\cal L}({\cal H}_Q)\bigr)$ to ${\cal L}({\cal H}_{RQ})$.
It takes a superoperator ${\cal S}$ on $Q$ to the operator
\begin{equation}
\bigl({\cal I}^R\otimes{\cal S}\bigr)\bigl(|\Psi\rangle\langle\Psi|\bigr)=
\sum_{j,k}|f_j\rangle\langle f_k|
\otimes
{\cal S}\bigl(|e_j\rangle\langle e_k|\bigr)
\;,
\label{OPdef}
\end{equation}
and we recover ${\cal S}$ via
\begin{equation}
\bigl\langle f_j,e_l\bigl|
\bigl({\cal I}^R\otimes{\cal S}\bigr)\bigl(|\Psi\rangle\langle\Psi|\bigr)
\bigr|f_k,e_m\bigr\rangle
=
\bigl\langle e_l\bigl|\,
{\cal S}\bigl(|e_j\rangle\langle e_k|\bigr)
\bigr|e_m\bigr\rangle
={\cal S}_{lj,mk}\;.
\end{equation}
Looking at OP in a slightly different way, 
\begin{equation}
\bigl({\cal I}^R\otimes{\cal S}\bigr)\bigl(|\Psi\rangle\langle\Psi|\bigr)=
\sum_{\alpha,\beta}
{\cal S}_{\alpha\beta}
|\Phi_{\tau_\alpha}\rangle\langle\Phi_{\tau_\beta}|\;,
\end{equation}
we find that
\begin{equation}
\bigl\langle\Phi_A\bigl|
\bigl({\cal I}^R\otimes{\cal S}\bigr)\bigl(|\Psi\rangle\langle\Psi|\bigr)
\bigr|\Phi_B\bigr\rangle=
(A|{\cal S}|B)\;.
\label{OPequiv}
\end{equation}
Thus the OP of a superoperator ${\cal S}$ operates in the same way as the
left-right action of ${\cal S}$.

We are ready now to return to complete positivity.  Recall that we are trying 
to show that any completely positive superoperator ${\cal A}$ has an operator 
decomposition.  The key point is that complete positivity requires that the 
OP of ${\cal A}$, i.e., 
$\bigl({\cal I}^R\otimes{\cal A}\bigr)\bigl(|\Psi\rangle\langle\Psi|\bigr)$,
be a positive operator, but Eq.~(\ref{OPequiv}) now shows this to be 
equivalent to the requirement that ${\cal A}$ be positive relative to its 
left-right action, which I write as ${\cal A}\ge0$.  Any such positive 
superoperator has many operator decompositions, including its orthogonal 
decomposition.  Moreover, we get for free, by using the decomposition theorem 
for positive operators \cite{Hughston1993a}, the following result, originally 
due to Choi \cite{Choi1975a}: two decompositions $A_\alpha$ and $B_\alpha$ 
give rise to the same completely positive superoperator if and only if they
are related by a unitary matrix $V_{\beta\alpha}$, i.e.,
\begin{equation}
B_\beta=\sum_\alpha V_{\beta\alpha}A_\alpha
\end{equation}
(if one decomposition has a smaller number of operators, it is extended
by appending zero operators).

What we have shown in this subsection is that a map is completely positive 
if and only if it is a positive superoperator relative to the left-right
action.  For a quantum operation we must add the trace-decreasing 
condition~(\ref{tdcond}), which now can be put in the compact form
\begin{equation}
{\cal A}^\cross(1)
=\sum_\alpha A_\alpha^\dagger A_\alpha
\le1\;,
\end{equation}
with equality if and only if the operation is trace preserving.  Thus we 
can now characterize a quantum operation as a superoperator that is positive 
relative to the left-right action (${\cal A}\ge0$, complete positivity) and 
that satisfies ${\cal A}^\cross(1)\le1$ (trace-decreasing). 

We should introduce one more ingredient before leaving our discussion of 
superoperators.  Suppose the initial state $\rho$ of $Q$ is VEC'ed to an 
initial joint state $|\Phi_{\!\sqrt\rho}\rangle$, as in Eq.~(\ref{vecpure}).  
The joint state of $RQ$ after the dynamics described by 
${\cal I}^R\otimes{\cal A}$ is given 
by 
\begin{equation}
\rho^{RQ'}=
{\bigl({\cal I}^R\otimes{\cal A}\bigr)
\bigl(
|\Phi_{\!\sqrt\rho}\rangle\langle\Phi_{\!\sqrt\rho}|
\bigr)
\over
\tr\bigl({\cal A}(\rho)\bigr)}
=
\biggl({\cal I}^R\otimes
{{\cal A}\circ\sqrt\rho\otimes\!\sqrt\rho\over\tr\bigl({\cal A}(\rho)\bigr)}
\biggr)
\!\bigl(|\Psi\rangle\langle\Psi|\bigr)\;.
\end{equation}
Referring to Eq.~(\ref{OPequiv}), we see that the superoperator
\begin{equation}
{\cal A}_\rho\equiv
{{\cal A}\circ\sqrt\rho\otimes\!\sqrt\rho\over\tr\bigl({\cal A}(\rho)\bigr)}
\label{Arho}
\end{equation}
is equivalent to the joint density operator $\rho^{RQ'}$; i.e., $\rho^{RQ'}$ 
operates the same as the left-right action of ${\cal A}_\rho$.

Straightforward consequences of the definition of ${\cal A}_\rho$ are 
that
\begin{eqnarray}
{\cal A}_\rho(1)
&=&{{\cal A}(\rho)\over\tr\bigl({\cal A}(\rho)\bigr)}
=\rho'\;,
\label{Arho1}\\
\sigma\equiv{\cal A}_\rho^\cross(1)
&=&{\sqrt\rho{\cal A}^\cross(1)\sqrt\rho\over\tr\bigl({\cal A}(\rho)\bigr)}
\le{\rho\over\tr\bigl({\cal A}(\rho)\bigr)}\;.
\label{Arhocross1}
\end{eqnarray}
If ${\cal A}$ is trace preserving, then the density operator $\sigma=\rho$.  
The physical significance of $\sigma$ can be ferreted out with a bit more 
work.  First we note that
\begin{equation}
\langle e_k|\sigma|e_j\rangle
=\tr\bigl({\cal A}_\rho^\cross(1)|e_j\rangle\langle e_k|\bigr)
=\tr\bigl({\cal A}_\rho(|e_j\rangle\langle e_k|)\bigr)\;.
\end{equation}
Writing the joint state of $RQ$ after the dynamics as
\begin{equation}
\rho^{RQ'}
=
\bigl({\cal I}^R\otimes{\cal A}_\rho\bigr)\bigl(|\Psi\rangle\langle\Psi|\bigr)
=
\sum_{j,k}
|f_j\rangle\langle f_k|\otimes
{\cal A}_\rho\bigl(|e_j\rangle\langle e_k|\bigr)\;,
\end{equation}
we find that the state of $R$ after the dynamics, 
\begin{equation}
\rho^{R'}=\tr_Q(\rho^{RQ'})=
\sum_{j,k}|f_j\rangle\langle f_k|\langle e_k|\sigma|e_j\rangle\;,
\end{equation}
is the ``transpose'' of $\sigma$ with respect to the bases $|e_j\rangle$
and $|f_j\rangle$.

\subsection{Reversal of quantum operations}

We are now ready to formulate the problem of reversing open-system 
dy\-na\-mics---i.e., correcting errors due to coupling to an environment.  
In this subsection I follow closely the formulation found in 
\cite{Nielsen1998a}.  The mathematical statement of the problem is the 
{\it reversibility\/} of a quantum operation ${\cal A}$ on a ``code 
subspace'' $C$ of the system Hilbert space ${\cal H}_Q$.  The reversal must 
be accomplished by a physical process, so it, too, is described by an 
operation, ${\cal R}$.  We want the reversal definitely to occur, so we 
require that ${\cal R}$ be a {\it trace-preserving\/} operation.  Thus we 
say that \cite{Schumacher1996b,Nielsen1997b,Nielsen1998a} {\it a quantum 
operation ${\cal A}$ is reversible on the code subspace $C$ if there exists 
a trace-preserving reversal operation ${\cal R}$, acting on the total 
state space of $Q$, such that} 
\begin{equation}
{\cal R}(\rho')
={{\cal R}\circ{\cal A}(\rho)\over\tr\bigl({\cal A}(\rho)\bigr)}
=\rho
\label{revcond}
\end{equation}
{\it for all $\rho$ whose support is confined to $C$}. An immediate consequence 
of the linearity of ${\cal R}\circ{\cal A}$ is that 
\begin{equation}
\tr\bigl({\cal A}(\rho)\bigr)=\mbox{constant}\equiv\mu^2
\label{mu}
\end{equation}
for all $\rho$ whose support is confined to $C$ \cite{Nielsen1997b}, where
$\mu$ is a real constant satisfying $0<\mu\le1$..  This allows us to rewrite 
the reversibility condition~(\ref{revcond}).  To do so, we introduce the 
restriction of ${\cal A}$ to $C$,
\begin{equation}
{\cal A}_C\equiv{\cal A}\circ
P_C\otimes P_C\;, 
\end{equation}
where $P_C$ is the projector onto $C$.  Then the reversibility condition
becomes
\begin{equation}
{\cal R}\circ{\cal A}_C=\mu^2 P_C\otimes P_C\;.
\end{equation}

It is not hard to show (for proofs see 
\cite{Bennett1996b,Knill1997a,Nielsen1998a}) that {\it a quantum operation 
${\cal A}$, with decomposition operators $A_\alpha$, is reversible 
on $C$ if and only if there exists a positive matrix $m_{\alpha\beta}$, 
having unit trace, such that}
\begin{equation}
P_CA_\beta^\dagger A_\alpha P_C=\mu^2m_{\alpha\beta}P_C\;.
\label{revcond2}
\end{equation}
It is instructive to rewrite this condition as 
$\langle e_j|A_\beta^\dagger A_\alpha|e_k\rangle
=\mu^2m_{\alpha\beta}\delta_{jk}$,
where the vectors $|e_j\rangle$ make up an orthonormal basis on $C$.
We can think of each decomposition operator $A_\alpha$ as an ``error
operator.''  Though each error operator acts like a multiple of a unitary 
operator within $C$, different error operators are not required to map 
to orthogonal subspaces.  How do we square this with the discussion 
in Sec.~\ref{errcorr}?

The puzzle is resolved by realizing that the decomposition of ${\cal A}$ is 
not unique.  We need to choose the decomposition operators to represent 
independent, indeed orthogonal errors within $C$.  To do so, take any 
density operator $\rho$ whose support is the entirety of $C$, and use a 
unitary matrix transformation of the decomposition operators $A_\alpha$ 
to diagonalize the matrix 
$\mu^2 m_{\alpha\beta}=\tr(A_\alpha\rho A_\beta^\dagger)$.  In the new
decomposition, called a {\it canonical decomposition} \cite{Nielsen1998a}, 
the decomposition operators $\tilde A_\alpha$ satisfy
\begin{equation}
(\tilde A_\beta\sqrt\rho\,|\tilde A_\alpha\sqrt\rho)
=\tr(\tilde A_\alpha\rho \tilde A_\beta^\dagger)
=\mu^2 \tilde m_{\alpha\beta}=\mu^2\lambda_\alpha\delta_{\alpha\beta}\;,
\label{cancond}
\end{equation}
where the eigenvalues $\lambda_\alpha$ satisfy $1=\sum_\alpha\lambda_\alpha$.  
In terms of this canonical decomposition, the reversal 
condition~(\ref{revcond2}) becomes
\begin{equation}
P_C\tilde A_\beta^\dagger \tilde A_\alpha P_C
=\mu^2\lambda_\alpha\delta_{\alpha\beta}P_C\;.
\label{revcond3}
\end{equation}

This condition has a ready interpretation.  When $\alpha=\beta$, the operator 
polar-de\-com\-po\-si\-tion theorem \cite{Peres1993a} implies that there 
exists a unitary operator $U_\alpha$ such that
\begin{equation}
\tilde A_\alpha P_C
=U_\alpha\sqrt{P_C\tilde A_\alpha^\dagger A_\alpha P_C}
=\mu\sqrt{\lambda_\alpha} U_\alpha P_C
=\mu\sqrt{\lambda_\alpha} P_\alpha U_\alpha\;.
\end{equation}
In the last equality we introduce the projector 
$P_\alpha\equiv U_\alpha P_C U_\alpha^\dagger$ onto the subspace that is 
the unitary image of $C$ under $U_\alpha$.  In terms of the canonical
decomposition the restriction of ${\cal A}$ to $C$ becomes
\begin{equation}
{\cal A}_C
=\sum_\alpha \tilde A_\alpha P_C\otimes P_C\tilde A_\alpha^\dagger
=\mu^2\sum_\alpha\lambda_\alpha 
U_\alpha P_C\otimes P_C U_\alpha^\dagger
=\mu^2\sum_\alpha\lambda_\alpha 
P_\alpha U_\alpha \otimes U_\alpha^\dagger P_\alpha\;.
\label{revA}
\end{equation}
When one or more of the eigenvalues $\lambda_\alpha$ is zero, we are 
dealing with a {\it degenerate code} \cite{Gottesman1996a,Nielsen1998a}:  
as far as the action of ${\cal A}$ within $C$ is concerned, the 
$\lambda_\alpha=0$ operators are irrelevant.   We discard those operators
henceforth, remembering that this is legitimate so long as we restrict 
attention to the code subspace.  The irrelevant decomposition operators 
discarded, the content of the $\alpha\ne\beta$ terms in Eq.~(\ref{revcond2}) 
is that the image subspaces $P_\alpha$ are orthogonal, i.e.,
\begin{equation}
P_\alpha P_\beta=\delta_{\alpha\beta}P_\beta\;.
\end{equation}
Thus in the canonical decomposition, the error operators $\tilde A_\alpha$ 
act like multiples of unitary operators on $C$ and map $C$ to orthogonal 
subspaces.

It is important for the considerations in Sec.~\ref{bounds} to note
that when we go to a canonical decomposition, we are diagonalizing the 
completely positive superoperator~(\ref{Arho}),
\begin{equation}
{\cal A}_\rho
={1\over\mu^2}\sum_\alpha
\tilde A_\alpha\sqrt\rho\otimes\!\sqrt\rho\tilde A_\alpha^\dagger
=\sum_\alpha
\lambda_\alpha U_\alpha\sqrt\rho\otimes\!\sqrt\rho\,U_\alpha^\dagger\
=\sum_\alpha
\lambda_\alpha 
\sqrt{\rho_\alpha}\,U_\alpha
\otimes 
U_\alpha^\dagger\sqrt{\rho_\alpha}\;,
\label{Arhodiag}
\end{equation}
for the operators $\tilde A_\alpha\sqrt\rho$ are orthogonal according to
Eq.~(\ref{cancond}).  In the last equality of Eq.~(\ref{Arhodiag}), we 
introduce the orthogonal density operators 
$\rho_\alpha=U_\alpha\rho\,U_\alpha^\dagger$ that $\rho$ is mapped to by 
the unitaries $U_\alpha$.  Notice that the eigenvalues and normalized 
eigenoperators of ${\cal A}_\rho$ (relative to the left-right action) are 
$\lambda_\alpha$ and $U_\alpha\sqrt\rho=\sqrt{\rho_\alpha}U_\alpha$.  The 
equivalence between ${\cal A}_\rho$ and $\rho^{RQ'}$ means that OP'ing the 
operators $U_\alpha\sqrt\rho$ generates the eigenvectors of $\rho^{RQ'}$.

For our purposes the only relevant part of the reversal operation is its 
restriction to the subspace $N$ that is the direct sum of the unitary images 
of $C$ under the unitaries $U_\alpha$.  This restriction, which is unique 
\cite{Nielsen1998a}, is given by
\begin{equation}
{\cal R}_N=
\sum_\alpha
U_\alpha^\dagger P_\alpha\otimes P_\alpha U_\alpha\;.
\end{equation}
It is easy to verify that this ${\cal R}_N$ reverses ${\cal A}_C$, and it 
also satisfies ${\cal R}_N^\cross(1)=\sum_\alpha P_\alpha=P_N$, the 
appropriate trace-preserving condition for the restriction.  It is 
interesting to note that since 
\begin{equation}
\rho'=
\sum_\alpha\lambda_\alpha U_\alpha\rho\,U_\alpha^\dagger
=\sum_\alpha\lambda_\alpha\rho_\alpha
\label{rhoprime}
\end{equation}
can be written in terms of an ensemble of orthogonal density operators
$\rho_\alpha$, we can give a very compact equation for the relevant part 
of the reversal operation:
\begin{equation}
{\cal R}_{\rho'}={\cal A}_\rho^\cross\;.
\end{equation}

\section{Entropy Exchange and Entanglement Fidelity} 
\label{bounds}

\subsection{Information-theoretic formulation of reversibility}

Our starting point in this section is Eq.~(\ref{rhoprime}), which gives
$\rho'$ in terms of an ensemble of orthogonal density operators $\rho_\alpha$.
This allows to conclude that
\begin{equation}
S(\rho')=
S(\rho)-\sum_\alpha\lambda_\alpha\log\lambda_\alpha
\;,
\label{Srhoprime}
\end{equation}
where $S(\rho)=-\tr(\rho\log\rho)$ is the von Neumann entropy of $\rho$.
Since the quantities $\lambda_\alpha$ are the eigenvalues of ${\cal A}_\rho$, 
the second term on the right is the entropy of the completely positive 
superoperator ${\cal A}_\rho$ or, equivalently, the entropy of $\rho^{RQ'}$.   
Schumacher \cite{Schumacher1996a} introduced this entropy and dubbed it 
the {\it entropy exchange}
\begin{equation}
S_e(\rho,{\cal A})\equiv
S(\rho^{RQ'})=
S({\cal A}_\rho)=
-\sum_\alpha\lambda_\alpha\log\lambda_\alpha\;.
\label{entex}
\end{equation}

It is useful at this point to introduce the {\it superoperator trace}
\begin{equation} 
\Tr({\cal S})  
\equiv
\sum_\alpha(\tau_\alpha|\,{\cal S}|\tau_\alpha)
=\tr\bigl(({\cal I}^R\otimes{\cal S})(|\Psi\rangle\langle\Psi|)\bigr)
=\tr\bigl({\cal S}(1)\bigr)
\;,
\end{equation}   
where the operators $\tau_\alpha$ make up an orthonormal operator basis and
where we use Eqs.~(\ref{OPdef}) and (\ref{OPequiv}) to reduce the definition
to an operator trace.  Not surprisingly, ${\cal A}_\rho$ has unit trace:
\begin{equation} 
\Tr\bigl({\cal A}_\rho\bigr)=
\tr\bigl({\cal A}_\rho(1)\bigr)=
\tr(\rho')=1\;.  
\end{equation}
The entropy exchange~(\ref{entex}) can now be written as
\begin{equation}
S({\cal A}_\rho)=-\Tr({\cal A}_\rho\log{\cal A}_\rho)\;.
\end{equation}

Equations~(\ref{mu}) and (\ref{Srhoprime}) are consequences of the 
reversibility of ${\cal A}$ on $C$.  It turns out that they are also 
sufficient to ensure reversibility \cite{Nielsen1998a}. {\it A quantum 
operation ${\cal A}$ is reversible on the code subspace $C$ if and only 
if the following two conditions are satisfied:
\vskip2pt
\noindent
Condition~1.
\begin{equation}
\tr\bigl({\cal A}(\rho)\bigr)=\mu^2
\label{cond1}
\end{equation}
for all $\rho$ whose support is confined to $C$, where $\mu$ is a real 
constant satisfying $0<\mu \le1$};
\vskip2pt
\noindent
{\it Condition~2.
\begin{equation}
S(\rho)=
S(\rho')-S({\cal A}_\rho)
\end{equation}
for any one $\rho$ whose support is the entirety of $C$ (and then for all 
$\rho$ whose support is confined to $C$)}.  A proof of this theorem is given 
in \cite{Nielsen1998a}.  Here I give a very simple proof of sufficiency, 
necessity already having been demonstrated.  

We start with the orthogonal decomposition of 
\begin{equation}
{\cal A}_\rho=
\sum_\alpha\lambda_\alpha\tau_\alpha\otimes\tau_\alpha^\dagger\;,
\label{Arhodecomp}
\end{equation}
where the eigenoperators are orthonormal, i.e., 
$(\tau_\alpha|\tau_\beta)=\delta_{\alpha\beta}$.  The polar-decomposition
theorem \cite{Peres1993a} guarantees that there exists a unitary operator
$U_\alpha$ such that
\begin{equation}
\tau_\alpha=U_\alpha\sqrt{\sigma_\alpha}=\sqrt{\rho_\alpha}\,U_\alpha\;,
\label{poldecomp}
\end{equation}
where $\sigma_\alpha=\tau_\alpha^\dagger\tau_\alpha$ and
$\rho_\alpha=\tau_\alpha\tau_\alpha^\dagger$ are normalized density
operators that are unitarily equivalent, i.e., 
$\rho_\alpha=U_\alpha\sigma_\alpha U_\alpha^\dagger$.  The unitary 
equivalence implies that $S(\rho_\alpha)=S(\sigma_\alpha)$.  
We have from Eqs.~(\ref{Arho1}) and (\ref{Arhocross1}) that 
\begin{eqnarray}
\rho'={\cal A}_\rho(1)&=&
\sum_\alpha\lambda_\alpha\rho_\alpha\;,\\
\sigma={\cal A}_\rho^\cross(1)&=&
\sum_\alpha\lambda_\alpha\sigma_\alpha
\;.
\end{eqnarray}
These ensembles for $\rho'$ and $\sigma$ give several inequalities
\cite{Wehrl1978a,Caves1994a}: 
\begin{eqnarray}
S({\cal A}_\rho)
&\ge&S(\rho')-\sum_\alpha\lambda_\alpha S(\rho_\alpha)
\ge 0\;,
\label{ineqSrhoprime}\\
S({\cal A}_\rho)
&\ge&S(\sigma)-\sum_\alpha\lambda_\alpha S(\sigma_\alpha)
\ge 0\;.
\label{ineqSsigma}
\end{eqnarray}
Equality holds on the left if and only if the density operators 
$\rho_\alpha$ ($\sigma_\alpha$) are orthogonal, whereas equality holds on 
the right if and only if the density operators $\rho_\alpha=\rho'$
($\sigma_\alpha=\sigma$) for all $\alpha$.  

We have not yet used the two Conditions.  If write Condition~1 as 
$\mu^2=\tr\bigl({\cal A}(\rho)\bigr)=\tr\bigl({\cal A}^\cross(1)\rho\bigr)$
for all $\rho$ whose support lies in $C$, we see that 
$P_C{\cal A}^\cross(1)P_C=\mu^2 P_C$, which implies
\begin{equation}
\sigma={\cal A}_\rho^\cross(1)=\rho\;.
\end{equation} 
Stringing together the left inequality in Eq.~(\ref{ineqSrhoprime}) and
the right inequality in Eq.~(\ref{ineqSsigma}) gives
\begin{equation}
S({\cal A}_\rho)\ge
S(\rho')-\sum_\alpha\lambda_\alpha S(\rho_\alpha)
=
S(\rho')-\sum_\alpha\lambda_\alpha S(\sigma_\alpha)\ge
S(\rho')-S(\rho)\;.
\end{equation}
Condition~2 dictates equality all the way across here: equality on the
left implies that the density operators $\rho_\alpha$ are orthogonal, and 
equality on the right implies that the density operators $\sigma_\alpha$ are 
all equal to $\rho$.  Thus we have that ${\cal A}_\rho$ has the 
form~(\ref{Arhodiag}), with orthogonal density operators $\rho_\alpha$.
Since $\rho$ has support on the entirety of $C$, we can put ${\cal A}_C$ in 
the reversible form~(\ref{revA}), thus completing the proof.

\subsection{Bounds on entropy exchange and entanglement fidelity}
\label{thebounds}

Thus far we have dealt with exact exact error correction.  Both more 
difficult and more important is approximate error correction, for which
we need a measure of the fidelity of a reversal.  Schumacher 
\cite{Schumacher1996a} has introduced a suitable measure,
\begin{equation}
F_e(\rho,{\cal A})\equiv
\bigl\langle\Phi_{\sqrt\rho}\bigl|\rho^{RQ'}\bigr|\Phi_{\sqrt\rho}\bigr\rangle
\;,
\end{equation}
called the {\it entanglement fidelity\/}; it measures the fidelity with 
which ${\cal A}$ preserves the entanglement of the primary system with the 
reference system.  Using the superoperator formalism of Sec.~\ref{supops}, 
we can write the entanglement fidelity as a superoperator matrix element:
\begin{equation}
F_e(\rho,{\cal A})=
(\sqrt\rho\,|{\cal A}_\rho|\sqrt\rho\,)
=(\rho|{\cal A}|\,\rho)
\;.
\end{equation}
Since ${\cal A}_\rho$ is positive and has unit trace relative to the 
left-right action, $F_e(\rho,{\cal A})=1$ if and only if $\sqrt\rho$ is a 
(normalized) eigenoperator of ${\cal A}_\rho$ with eigenvalue 1, i.e., 
${\cal A}_\rho=\sqrt\rho\otimes\!\sqrt\rho$.  In this case ${\cal A}_\rho$ 
is a {\it pure\/} completely positive superoperator; both the von Neumann 
entropy (or entropy exchange), $S({\cal A}_\rho)=0$, and the {\it quadratic 
entropy\/}, $\Tr({\cal A}_\rho^2)=1$, faithfully report this purity.  To 
analyze approximate reversal, we need relations between these two measures 
of purity and the entanglement fidelity.  Obtaining such relations is 
the task of this subsection.

We look first at ways to use the purity measures to bound the entanglement
fidelity away from 1.  Let the operators $\eta_\alpha$, $\alpha=1,\ldots,D^2$, 
be an orthonormal operator basis chosen so that $\eta_1=\sqrt\rho$, which
means that $F_e=(\eta_1|{\cal A}_\rho|\eta_1)$.  Then we have
\begin{equation}
S({\cal A}_\rho)\le
-\sum_{\alpha=1}^{D^2}
(\eta_\alpha|{\cal A}_\rho|\eta_\alpha)
\log(\eta_\alpha|{\cal A}_\rho|\eta_\alpha)
=-F_e\log F_e
-\sum_{\alpha=2}^{D^2}
(\eta_\alpha|{\cal A}_\rho|\eta_\alpha)
\log(\eta_\alpha|{\cal A}_\rho|\eta_\alpha)
\;.
\label{1ststep}
\end{equation}
The inequality here is the standard result that the von Neumann entropy is
never greater than the entropy calculated in any orthonormal basis 
\cite{Wehrl1978a,Caves1994a}.  Renormalizing the distribution remaining in 
the sum, we can write
\begin{equation}
S({\cal A}_\rho)\le h(F_e)
+(1-F_e)
\Biggl(
-\sum_{\alpha=2}^{D^2}
{(\eta_\alpha|{\cal A}_\rho|\eta_\alpha)\over 1-F_e}
\log{(\eta_\alpha|{\cal A}_\rho|\eta_\alpha)\over 1-F_e}
\Biggr)
\;,
\label{2ndstep}
\end{equation}
where $h(x)\equiv-x\log x-(1-x)\log(1-x)$ is the binary entropy.  Using 
the fact that the entropy within the large parentheses is bounded above 
by $\log(D^2-1)$, we get Schumacher's \cite{Schumacher1996a} {\it quantum 
Fano inequality},
\begin{equation}
S({\cal A}_\rho)\le h(F_e)+(1-F_e)\log(D^2-1)\;,
\end{equation}
which bounds the entanglement fidelity away from one.  Schumacher obtained
the quantum Fano inequality by applying the same reasoning to $\rho^{RQ'}$.

We can do the same thing with the quadratic entropy.  The analogue of
Eqs.~(\ref{1ststep}) and (\ref{2ndstep}) is
\begin{equation}
\Tr({\cal A}_\rho^2)
=
\sum_{\alpha,\beta}
|(\eta_\alpha|{\cal A}_\rho|\eta_\beta)|^2
\ge
\sum_{\alpha=1}^{D^2}
(\eta_\alpha|{\cal A}_\rho|\eta_\alpha)^2
=
F_e^2+
(1-F_e)^2
\sum_{\alpha=2}^{D^2}
\Biggl({(\eta_\alpha|{\cal A}_\rho|\eta_\alpha)\over1-F_e}\Biggr)^{\!\!2}
\;.
\end{equation}
Since the remaining sum is bounded below by $(D^2-1)^{-1}$, we get a
{\it quadratic quantum Fano inequality},
\begin{equation}
\Tr({\cal A}_\rho^2)
\ge
F_e^2+
{(1-F_e)^2\over D^2-1}
\;,
\end{equation}
which like the entropy version, bounds the entanglement fidelity away from
one.

We now turn to the opposite task: using the purity measures to place lower
bounds on the entanglement fidelity.  That this might work is suggested by 
the case $S({\cal A}_\rho)=0$ or, equivalently, $\Tr({\cal A}_\rho^2)=1$.  
Then ${\cal A}_\rho=\tau\otimes\tau^\dagger$ is pure, the normalized 
eigenoperator $\tau$ having eigenvalue 1.  If ${\cal A}$ is trace preserving, 
we have that $\rho={\cal A}_\rho^\cross(1)=\tau^\dagger\tau$.  Then, by the 
polar-decomposition theorem, there exists a unitary operator $U$ such that 
$\tau=U\!\sqrt\rho$.  Defining a new operation by 
\begin{equation}
{\cal A}'\equiv U^\dagger\otimes U\circ{\cal A}\;,
\label{Arhoprime}
\end{equation}
we have that ${\cal A}'_\rho=\sqrt\rho\otimes\!\sqrt\rho$, so that 
$F_e(\rho,{\cal A}')=1$.  What we have shown is that ${\cal A}$ is within
a unitary of an operation that has unity entanglement fidelity for input
density operator $\rho$.

We now mimic this construction when ${\cal A}_\rho$ is not pure.  In doing 
so, we {\it assume that ${\cal A}$ is trace preserving}.  Suppose that 
${\cal A}_\rho$ has the eigenvalue decomposition~(\ref{Arhodecomp}). Let 
$\lambda_1$ be the largest eigenvalue, and let 
\begin{equation}
\tau_1=U\sqrt{\tau_1^\dagger\tau_1}=U\sqrt\sigma_1\;,
\end{equation} 
as in Eq.~(\ref{poldecomp}).  Define the new operation~(\ref{Arhoprime}), 
with the result that
\begin{equation}
{\cal A}'_\rho
=\lambda_1\sqrt{\sigma_1}\otimes\!\sqrt{\sigma_1}+
\sum_{\alpha\ne1}
\lambda_\alpha U^\dagger\tau_\alpha\otimes\tau_\alpha^\dagger U
\;.
\end{equation}
Now we use the fact that ${\cal A}$ is trace preserving to write
\begin{equation}
\rho={\cal A}_\rho^\cross(1)=
\lambda_1\sigma_1+
\sum_{\alpha\ne1}\lambda_\alpha\tau_\alpha^\dagger\tau_\alpha
\;,
\end{equation}
which implies that $\rho\ge\lambda_1\sigma_1$.  Because the operator 
square-root function is an {\it operator-monotone function\/} 
\cite{Marshall1979a}, this operator inequality remains true upon taking
the square root of both sides:
\begin{equation}
\sqrt\rho\ge\sqrt{\lambda_1}\sqrt{\sigma_1}\;.
\label{rootineq}
\end{equation}  
The operator-monotone property of the square root is  is proved in the 
Appendix.  By writing $\sigma_1$ in terms of its eigendecomposition, we 
see that the operator inequality~(\ref{rootineq}) implies that
\begin{equation}
\tr(\sqrt\rho\sqrt{\sigma_1})\ge 
\sqrt{\lambda_1}\tr(\sigma_1)=
\sqrt{\lambda_1}
\;.
\end{equation}
Now we notice that 
\begin{equation}
F_e(\rho,{\cal A}')=
(\sqrt\rho\,|{\cal A}'_\rho|\sqrt\rho\,)
\ge
\lambda_1\bigl(\tr(\sqrt\rho\sqrt{\sigma_1})\bigr)^2
\ge\lambda_1^2
\;.
\label{FeAprime}
\end{equation}
This is our key result.  It says that if the largest eigenvalue of 
${\cal A}_\rho$ is close to 1, then ${\cal A}$ can be corrected by a 
unitary so that the entanglement fidelity is close to 1.

We can translate Eq.~(\ref{FeAprime}) into weaker bounds that involve
the purity measures.  For the entropy exchange we have that 
\begin{equation}
S({\cal A}_\rho)=
-\sum_\alpha\lambda_\alpha\log\lambda_\alpha
\ge
-\sum_\alpha\lambda_\alpha\log\lambda_1
=-\log\lambda_1\;.
\end{equation}
Using Eq.~(\ref{FeAprime}), we obtain a {\it quantum anti-Fano inequality}
\begin{equation}
F_e(\rho,{\cal A}')\ge
\exp\Bigl(-2S({\cal A}_\rho)\Bigr)
\;.
\end{equation}
For the quadratic entropy we can write
\begin{equation}
\Tr({\cal A}_\rho^2)=
\sum_\alpha\lambda_\alpha^2
\le\sum_\alpha\lambda_\alpha\lambda_1
=\lambda_1
\;.
\end{equation}
Again using Eq.~(\ref{FeAprime}), we obtain a {\it quadratic quantum anti-Fano
inequality}
\begin{equation}
F_e(\rho,{\cal A}')\ge\Bigl(\Tr({\cal A}_\rho^2)\Bigr)^2
\;.
\end{equation}
Both of the anti-Fano inequalities place lower bounds on the entanglement 
fidelity as we had hoped.  It should be remembered that they apply only
to trace-preserving operations.

\appendix
\section*{Appendix}

Let $A$ and $B$ be positive operators such that $A\ge B$.  Define the 
Hermitian operator
\begin{equation}
F\equiv\sqrt A-\sqrt B=\sum_k F_k|f_k\rangle\langle f_k|\;,
\end{equation}
where the sum is the eigenvalue decomposition of $F$.  The eigenvalues
of $F$ satisfy
\begin{equation}
F_k=
\langle f_k|\sqrt A\,|f_k\rangle
-\langle f_k|\sqrt B\,|f_k\rangle
\ge
-\langle f_k|\sqrt B\,|f_k\rangle
\;.
\label{Fk1}
\end{equation}
Now define the Hermitian operator
\begin{equation}
G\equiv A-B=\sqrt B F+F\sqrt B+F^2
\end{equation}
Since $G\ge0$, we have
\begin{equation}
0\le\langle f_k|G|f_k\rangle
=2F_k\langle f_k|\sqrt B\,|f_k\rangle+F_k^2\;.
\label{Fk2}
\end{equation}
Equations~(\ref{Fk1}) and (\ref{Fk2}) together imply that $F_k\ge0$,
which means that $\sqrt A\ge\sqrt B$.

\newpage
\begin{table}[ht]
\caption{How the 27 single-qubit error operators affect an arbitrary
superposition of the logical zero $|\zero\rangle_L=|\up\up\up\rangle$
and the logical one $|\one\rangle_L=|\down\down\down\rangle$ in Shor's
nine-bit code.  The first line shows the superposition state with no error.}
\label{table1}
\vspace{25pt}
\begin{tabular*}{6.5in}{l@{\extracolsep{\fill}}cc}
Error&Error operator&State after error\\[6pt]
\hline
\\
no error&
$1\otimes1\otimes1\otimes1\otimes1\otimes1\otimes1\otimes1\otimes1$&
$\alpha|\up\up\up\rangle+\beta|\down\down\down\rangle$\\[6pt]
bit flip on 1st qubit&
$\sigma_1\otimes1\otimes1\otimes1\otimes1\otimes1\otimes1\otimes1\otimes1$&
$\alpha|\up_1\up\up\rangle+\beta|\down_1\down\down\rangle$\\
bit flip on 2nd qubit&
$1\otimes\sigma_1\otimes1\otimes1\otimes1\otimes1\otimes1\otimes1\otimes1$&
$\alpha|\up_2\up\up\rangle+\beta|\down_2\down\down\rangle$\\ 
bit flip on 3rd qubit&
$1\otimes1\otimes\sigma_1\otimes1\otimes1\otimes1\otimes1\otimes1\otimes1$&
$\alpha|\up_3\up\up\rangle+\beta|\down_3\down\down\rangle$\\[6pt]
bit flip on 4th qubit&
$1\otimes1\otimes1\otimes\sigma_1\otimes1\otimes1\otimes1\otimes1\otimes1$&
$\alpha|\up\up_1\up\rangle+\beta|\down\down_1\down\rangle$\\ 
bit flip on 5th qubit&
$1\otimes1\otimes1\otimes1\otimes\sigma_1\otimes1\otimes1\otimes1\otimes1$&
$\alpha|\up\up_2\up\rangle+\beta|\down\down_2\down\rangle$\\ 
bit flip on 6th qubit&
$1\otimes1\otimes1\otimes1\otimes1\otimes\sigma_1\otimes1\otimes1\otimes1$&
$\alpha|\up\up_3\up\rangle+\beta|\down\down_3\down\rangle$\\[6pt] 
bit flip on 7th qubit&
$1\otimes1\otimes1\otimes1\otimes1\otimes1\otimes\sigma_1\otimes1\otimes1$&
$\alpha|\up\up\up_1\rangle+\beta|\down\down\down_1\rangle$\\ 
bit flip on 8th qubit&
$1\otimes1\otimes1\otimes1\otimes1\otimes1\otimes1\otimes\sigma_1\otimes1$&
$\alpha|\up\up\up_2\rangle+\beta|\down\down\down_2\rangle$\\ 
bit flip on 9th qubit&
$1\otimes1\otimes1\otimes1\otimes1\otimes1\otimes1\otimes1\otimes\sigma_1$&
$\alpha|\up\up\up_3\rangle+\beta|\down\down\down_3\rangle$\\[6pt] 
phase flip on 1st qubit&
$\sigma_3\otimes1\otimes1\otimes1\otimes1\otimes1\otimes1\otimes1\otimes1$&
\\ 
phase flip on 2nd qubit&
$1\otimes\sigma_3\otimes1\otimes1\otimes1\otimes1\otimes1\otimes1\otimes1$&
$\alpha|\down\up\up\rangle+\beta|\up\down\down\rangle$\\ 
phase flip on 3rd qubit&
$1\otimes1\otimes\sigma_3\otimes1\otimes1\otimes1\otimes1\otimes1\otimes1$&
\\[6pt] 
phase flip on 4th qubit&
$1\otimes1\otimes1\otimes\sigma_3\otimes1\otimes1\otimes1\otimes1\otimes1$&
\\ 
phase flip on 5th qubit&
$1\otimes1\otimes1\otimes1\otimes\sigma_3\otimes1\otimes1\otimes1\otimes1$&
$\alpha|\up\down\up\rangle+\beta|\down\up\down\rangle$\\ 
phase flip on 6th qubit&
$1\otimes1\otimes1\otimes1\otimes1\otimes\sigma_3\otimes1\otimes1\otimes1$&
\\[6pt] 
phase flip on 7th qubit&
$1\otimes1\otimes1\otimes1\otimes1\otimes1\otimes\sigma_3\otimes1\otimes1$&
\\ 
phase flip on 8th qubit&
$1\otimes1\otimes1\otimes1\otimes1\otimes1\otimes1\otimes\sigma_3\otimes1$&
$\alpha|\up\up\down\rangle+\beta|\down\down\up\rangle$\\ 
phase flip on 9th qubit&
$1\otimes1\otimes1\otimes1\otimes1\otimes1\otimes1\otimes1\otimes\sigma_3$
\\[6pt] 
phase-bit flip on 1st qubit& 
$-i\sigma_2\otimes1\otimes1\otimes1\otimes1\otimes1\otimes1\otimes1\otimes1$&
$\alpha|\down_1\up\up\rangle+\beta|\up_1\down\down\rangle$\\ 
phase-bit flip on 2nd qubit&
$1\otimes-i\sigma_2\otimes1\otimes1\otimes1\otimes1\otimes1\otimes1\otimes1$&
$\alpha|\down_2\up\up\rangle+\beta|\up_2\down\down\rangle$\\ 
phase-bit flip on 3rd qubit&
$1\otimes1\otimes-i\sigma_2\otimes1\otimes1\otimes1\otimes1\otimes1\otimes1$&
$\alpha|\down_3\up\up\rangle+\beta|\up_3\down\down\rangle$\\[6pt] 
phase-bit flip on 4th qubit&
$1\otimes1\otimes1\otimes-i\sigma_2\otimes1\otimes1\otimes1\otimes1\otimes1$&
$\alpha|\up\down_1\up\rangle+\beta|\down\up_1\down\rangle$\\ 
phase-bit flip on 5th qubit&
$1\otimes1\otimes1\otimes1\otimes-i\sigma_2\otimes1\otimes1\otimes1\otimes1$&
$\alpha|\up\down_2\up\rangle+\beta|\down\up_2\down\rangle$\\ 
phase-bit flip on 6th qubit&
$1\otimes1\otimes1\otimes1\otimes1\otimes-i\sigma_2\otimes1\otimes1\otimes1$&
$\alpha|\up\down_3\up\rangle+\beta|\down\up_3\down\rangle$\\[6pt] 
phase-bit flip on 7th qubit&
$1\otimes1\otimes1\otimes1\otimes1\otimes1\otimes-i\sigma_2\otimes1\otimes1$&
$\alpha|\up\up\down_1\rangle+\beta|\down\down\up_1\rangle$\\ 
phase-bit flip on 8th qubit&
$1\otimes1\otimes1\otimes1\otimes1\otimes1\otimes1\otimes-i\sigma_2\otimes1$&
$\alpha|\up\up\down_2\rangle+\beta|\down\down\up_2\rangle$\\ 
phase-bit flip on 9th qubit&
$1\otimes1\otimes1\otimes1\otimes1\otimes1\otimes1\otimes1\otimes-i\sigma_2$&
$\alpha|\up\up\down_3\rangle+\beta|\down\down\up_3\rangle$\\
\end{tabular*}
\end{table}

\end{document}